\documentclass[showpacs,preprintnumbers,showkeys,twocolumn,prb,floatfix]{revtex4}
\usepackage{amsmath}
\usepackage{epsfig}

\newcommand{\figWidth}{8cm}
\newcommand{\inThesis}[1]{}
\newcommand{\inPaper}[1]{#1}
\newcommand{\Ham}{{\cal H}}
\newcommand{\bx}{{\bf x}}
\newcommand{\bk}{{\bf k}}
\newcommand{\by}{{\bf y}}
\newcommand{\bn}{{\bf n}}
\newcommand{\bu}{{\bf u}}
\newcommand{\bheps}{{\bf\hat\epsilon}}
\newcommand{\bhrho}{{\bf\hat\rho}}
\newcommand{\bD}{{\bf\Delta}}

\begin{document}

\title{Phonon overlaps, polaron overlaps, and the effects of molecular size
on quantum tunneling in polyacetylene}
\date{\today}

\author{Connie Te-ching Chang and James P. Sethna}
\affiliation{Laboratory of Atomic and Solid State Physics (LASSP), Clark Hall,
Cornell University, Ithaca, NY 14853-2501, USA}
\date{\today}

\begin{abstract}
We provide a theory for the effects of polarons and phonons in mediating
and suppressing the quantum tunneling of electrons into single molecules of
conducting polymers, motivated by experiments on molecular quantum dots.  The
effects of both phonons and excitations of the polaron particle-in-a-box
states are calculated.\inThesis{
}
Using the Su-Schrieffer-Heeger (SSH) model of polyacetylene, we calculate the 
suppression of
ground--state to ground--state transitions and the position and strength of the
phonon and polaron side-bands.  
\end{abstract}

 
\maketitle

\section{Introduction}
Molecular quantum dots\cite{ho,stipe,smit,park1,park2} and conduction along one-dimensional molecules
(particularly carbon nanotubes\cite{tans,bockrath}) have both attracted much attention in the 
nanophysics experimental community.  Electron tunneling into molecular quantum dots is
known to occur together with the emission of particular phonon modes.  The
relative rates of emitting different phonons has been calculated 
 \inThesis{the previous chapter}\inPaper{\cite{C72}} using overlap integrals of
 the phonon wavefunction.  Tunneling into one-dimensional polymeric molecules
 will have an additional complication: the electron final state will be a
 polaron, dressed by a local screening cloud of phonons.  Polaronic effects,
 although likely small for stiff carbon nanotubes, are important in other
 one-dimensional conducting polymers.

\inThesis{In this chapter}\inPaper{Here}, we calculate the suppression of
tunneling due to polaron formation in the conducting polymer polyacetylene,
using the Su-Schrieffer-Heeger (SSH) model~\cite{SSH}.
We calculate the phonon sidebands directly and
provide the theory for calculating also the excitations into polaron-in-a-box
excited states which should form a characteristic size-dependent series of
excitation lines.  

\section{Polyacetylene}
Polyacetylene\cite{SSH2} is an organic, quasi one-dimensional semi-conducting polymer.  
Many experiments have been performed to characterize the physical properties of
polyacetylene such as its band gap, dimerization constant, lattice spacing,
behavior under doping, and crystal structure (packing).
In its stable, dimerized form, {\it{trans}}-polyacetylene has a backbone
consisting of alternating single and double bonds between adjacent CH groups.  This
structure and the degenerate ground states resulting 
from changing the order of
single and double bonding within the molecule accounts for many 
of the unusual properties of {\it{trans}}-polyacetylene 
such as the formation of solitons.  Here, because we add only one electron, the
ground state will form an ordinary polaron rather than solitons.  Hence, 
the unusual properties of
polyacetylene are not central to our analysis: the polaronic effects we examine
will be shared by more mundane conducting polymers.    
\subsection{The Hamiltonian}
\label{sec:Hamiltonians}
Polyacetylene can be simply modeled by focusing on the behavior of the free
carbon valence electron occupying the $p_{Z}$ orbital.  The
semi-conducting properties are thus due to polyacetylene's half-filled
$\pi$ band.

Following Su, Schrieffer and Heeger\cite{SSH,SSH2}, the
electronic band structure can be
calculated using the tight-binding approximation for the electrons in the
$p_{z}$ orbitals. The unit cell will have a basis consisting of
two CH groups with one
$p_{z}$ electron per group.  \inThesis{ Thus, there will be two
bands in the first
Brillouin zone -- the valence band and the conduction band,
and only the valence band will be filled.}  The simple model is valid where the
$\pi$-band width is large compared to the band gap, and where the short-range
electron-electron interactions are small compared to
the valence band width.

The SSH Hamiltonian for the lattice model of polyacetylene is:
  \begin{equation}  
  \Ham=\Ham_{el}+\Ham_{el-ph}+\Ham_{ph}
  \end{equation}
where there are contributions from the electrons in the $\pi$-band, the
$\pi$-band electron interaction with phonons, and the phonons.

The tight-binding Hamiltonian written in bra-ket notation gives the electronic contribution:
\begin{equation}
\label{eq:first}
\Ham_{el}=  
-\sum_{n,s} t_0[|n+1,s \rangle\langle n,s|+|n,s\rangle\langle n+1,s|]
\end{equation}
where $t_{0}$ is the hopping matrix element connecting nearest neighbors,
spin $s=\pm{1\over{2}}$, and $n$ denotes the index of the CH group for which 
$|n,\pm{1\over{2}}\rangle$ are the $p_Z$ orbital states.
The hopping 
matrix element in the absence of dimerization is represented by $t_0$.

The electron-phonon interaction is:
\begin{eqnarray}
\Ham_{el-ph}&=&\alpha\sum_{n,s} (u_{n+1}-u_{n})(|n+1,s \rangle\langle
n,s|\cr
&+&|n,s\rangle\langle n+1,s|)
\end{eqnarray}
where $\alpha$ is the electron-phonon coupling constant and the electron-phonon
coupling is expanded to first order about the undimerized state assuming a 
small dimerization.

The lattice component is given by:
\begin{equation} 
\label{eq:second}
\Ham_{ph}=
{K\over{2}}\sum_{n}(u_{n+1}-u_{n})^{2}+{M\over{2}}\sum_{n}\dot{u}_{n}^{2}
\end{equation} 
where the lattice coordinate $u_{n}$ defines the horizontal displacement of the
$n^{th}$ CH group away from its undisplaced lattice position at $na$ with $a$
the characteristic lattice length.  $M$ is the mass of the CH group and $K$ is
the spring constant for the $\sigma$-bonding backbone.  For a 
dimerized chain, $u_{n}=(-1)^{n}u_{0}$, where $u_{0}$ is 
the dimerization constant.  
The constants $t_{0}=2.5 eV$, $K=21 eV/$\AA, and $a=1.22$\AA\ were 
chosen following Su, Schrieffer, and Heeger~\cite{SSH}.
\inThesis{
\begin{figure}
\begin{center}
\leavevmode
\includegraphics[width=\figWidth]{polyFig/energy.eps}
\caption{Energy for neutral $C_{20}H_{22}$}
\end{center}
\label{Evsu0}
\end{figure}
}
Writing down the Bloch wavefunctions for the band states and 
diagonalizing the electron Hamiltonian yields an 
expression for the energy in terms of the crystal wave vector $\bk$, the 
dimerization $u$, 
and the constants above.  As the number of CH groups, $N$, tends to 
infinity, the sum in the expression for the electronic energy can 
be transformed into an integral; we solve it here for finite $N$.  
The SSH model spontaneously dimerizes to lower its energy, with $u_0\approx\pm0.04$\AA.
\inThesis{
Plotting energy per CH group vs. $u_{0}$, we see that the energy is minimized
for $u_{0}\approx\pm .04$\AA, reflecting the double degeneracy of the ground
state (figure~(3.1).}
\section{Finite Length Chains}   
\label{sec:Discrete Chains}
\begin{figure}
\begin{center}
\leavevmode
\includegraphics[width=\figWidth]{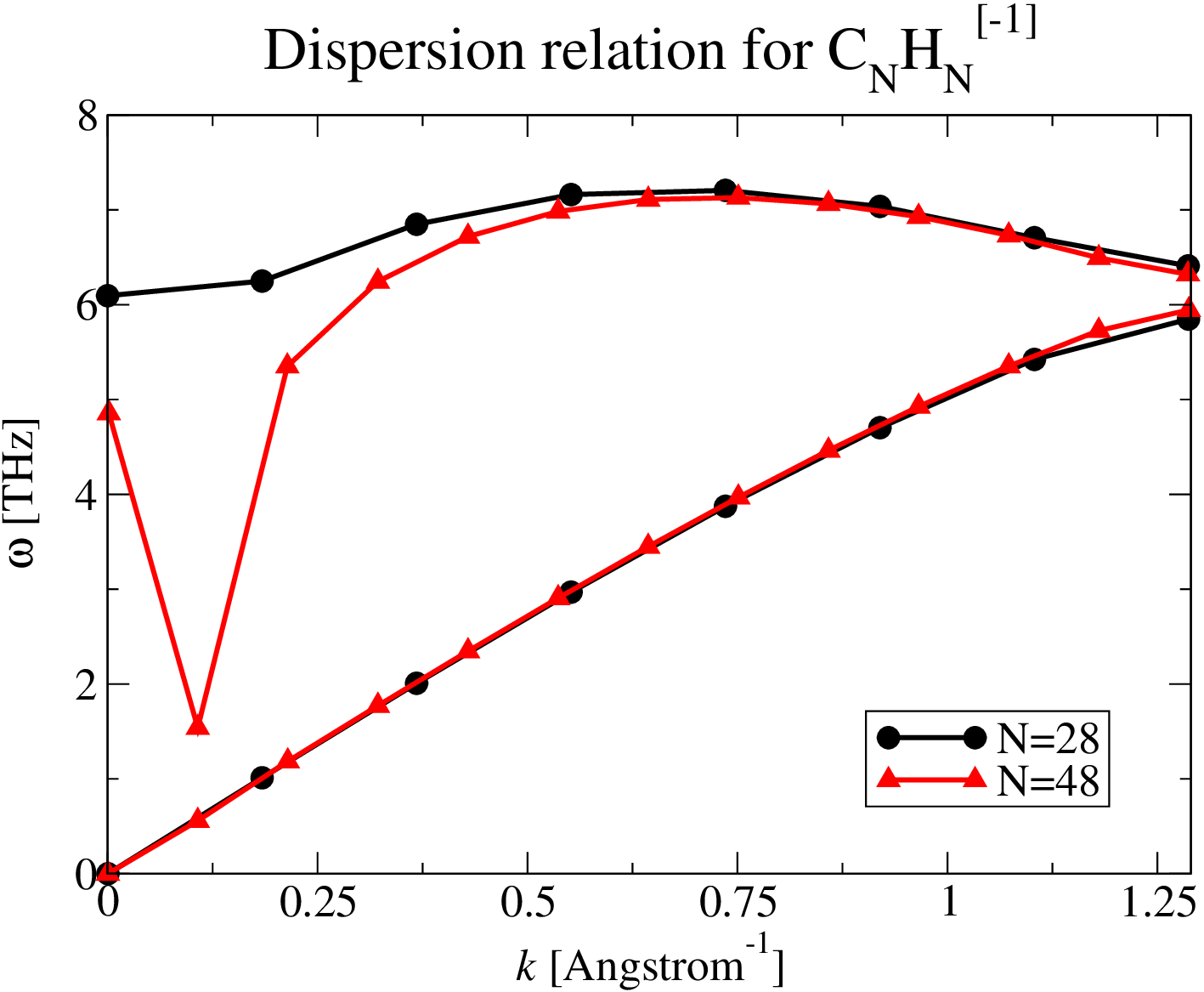}
\end{center}
   \caption{Dispersion relation for charged polyacetylene chains of length N}
\label{disp}
\end{figure}

Why are phonon sidebands seen for electronic transitions into molecular quantum
dots when they have not been important in the larger mesoscale dots?  
We address
this question by comparing the phonon overlaps for molecules of
different sizes/lengths.  To be specific, we consider finite
chains of polyacetylene 
with a discrete number $N$ of CH groups.  For simplicity, we
consider the case of periodic boundary conditions such that the first and 
last CH groups are
linked.  Given that we are looking at 
dimerized
polyacetylene chains, $N$ must be even to avoid imposing a soliton into 
the molecule.  In addition, the properties of small, circular SSH molecules
are known to have a sharp alternation between lengths $N=4n$ and $N=4n+2$; we
focus here on lengths which are multiples of four to avoid distraction by these
finite-size oscillations\inThesis{ (see figures~(\ref{EoverN}),~(\ref{minu0}),and~(3.2))}.  
From the hessian, we can extract a dispersion 
relation for
various lengths of chain.  Figure~(\ref{disp}) shows the dispersion 
relations for 
charged polyacetylene chains of length N=28 and N=48.  
Note that the range that the frequencies fall in are about the same for both
chains (0-32 tHz).  However, their dispersion 
relation differs. 
The acoustic branches (neighboring carbons moving in concert) are 
rather similar.  The optical branch (alternating displacement directions
for neighboring carbons) of the shorter chain exhibits
 a classic
dispersion relation for a dimerized system \cite{nakahara}, but the 
longer chain
shows a pronounced softening, foreshadowing the formation of a polaron for
chains of $N>50$.  
Notice in particular the softening of the optical branch at the mode
$\bk_{0}$ with one wavelength in the length of the chain; 
the polaron when it first forms near $N=52$ has a roughly sinusoidal 
form with this wavelength (see Fig. 3).  
%\subsection{Calculating Overlaps}
The zero-phonon emission line for an electronic transition is suppressed by the
square of the overlap integral $O_{0,0}$ between the initial and final 
vibrational ground states.  This is traditionally measured using the total $g$-factor:
\begin{equation}
\label{G-factor}
G=-\log(P_{00})=-\log[(O_{00})^2]
\end{equation}
where $P_{00}=e^{-G}$ is the probability that the molecule will 
remain in its ground
vibrational state with the addition of an electron.

The phonon sidebands and extra lines on the $dI/dV$ curves arise from overlaps
between the initial ground state and excited final states.  The integrated
strengths of these transitions (the sum of the heights for the
vibration-induced steps on the I-V curve) is given by $1-P_{00}$. Thus 
large $G$ is a
signature of the importance of phonon emission.  
We first focus on the dimerization mode,
changes of which dominate the lattice deformation for small $C_xH_{x}$ 
rings in the SSH model.  
\subsection{Dimerization mode only}
\label{oned}
We start by restricting our
attention purely to the change in dimerization $u_{0}$ in the SSH
model as the N-monomer system
changes from neutral to charged. To realize this simple picture, we treat the polyacetylene chain as a
one-variable problem, optimizing only the dimerization
constant $u_{0}$.  We find the minimum $u_{0}$ for neutral and charged chains of various lengths $N$
by scanning the energy over different values of $u_{0}$ and constraining the
displacements of CH units to be of the form $u_{n}=(-1)^{n}u_{0}$.  

Specifically, the dimerization of the initial
neutral state $u_0^{(1)}$ has an associated net Euclidean displacement 
distance 
in $N$-dimensional configuration space of 
$r_1=u_0^{(1)}\sqrt{N}$.  Similarly,
$r_2=u_0^{(2)}\sqrt{N}$ for the charged final state, so the net displacement is
$d=(u_0^{(2)}-u_0^{(1)})\sqrt{N}$ for our oscillator ($N$ orthogonal coordinates
all shifting by $u_0^{(2)}-u_0^{(1)}$).  The frequencies are given by
the optical phonon frequency (where neighboring CH units move in opposing directions) in the neutral ($\omega_1$) and charged 
($\omega_2$) states.  The one-dimensional formula for the overlap is:
\begin{eqnarray}
O_{0,0}&=&\int{dx}\left({{m_{CH}\omega_1}\over{\pi\hbar}}\right)^{1/4}
e^{-{{m\omega_1(x-r_1)^2}\over{2\hbar}}}\cr
&\times &\left({{m_{CH}\omega_2}\over{\pi\hbar}}\right)^{1/4}
e^{-{{m\omega_2(x-r_2)^2}\over{2\hbar}}}\cr
&=&\left({{\omega_1
    \omega_2}\over{\left({{\omega_1+\omega_2}\over{2}}\right)^2}}\right)^{1/4}
e^{-{{md^2}\over{2\hbar}}\left({{\omega_1 \omega_2}
\over{\omega_1+\omega_2}}\right)}.
\end{eqnarray}
where the displacement $d$ and the frequencies $\omega_1$ and $\omega_2$ are
defined above.  For example, this value of the overlap yields 
$O_{0,0}=0.468$ and $G=-\log|O_{0,0}|^2=1.52$ for a twelve unit polyacetylene chain.

Note that $G$
falls off as $1\over{N}$ for large $N$ in figure~(\ref{onemodecomp2}).  
In appendix A, we consider broadly the finite-size
effects of both optical phonons (considered in the above calculation) and
acoustic phonons.  We will see quite generally that optical phonon
overlaps for delocalized electrons go to one as $N\rightarrow\infty$
($G\approx1/N$), while acoustic phonons induce overlaps which go to
a constant for large $N$.  Hence, in the absence of polaronic distortions,
optical phonon emission becomes less important for larger molecules, 
perhaps explaining why phonon sidebands first became significant
when the dots became molecular.
%\begin{figure}[tbh]
%\begin{center}
%\leavevmode
%\includegraphics[width=\figWidth]{onemode.eps}
%\end{center}
%\caption{$g_{total}$ for 1-D case}
%\label{onedg}
%\end{figure}

%For comparison's sake, we also plot our data alongside a true $1\over{{N}}$
%    functional dependence to demonstrate that our plot follows the expected
%    form in figure~(\ref{onemodecomp}) where the circles denote our numbers and
%    the triangles denote the functional dep.
%\begin{figure}
%\begin{center}
%\leavevmode
%\includegraphics[width=\figWidth]{onemode_mod.eps}
%\end{center}
%\caption{N-dependence of $G$ in dimerization-only case where the $1/N$
%  dependence is clearly seen in this plot of $NG$ vs. $N$ since
%  $NG$ approaches a constant as $N$ increases.}
%\label{onemodecomp}
%\end{figure}
\begin{figure}
\begin{center}
\leavevmode
\includegraphics[width=\figWidth]{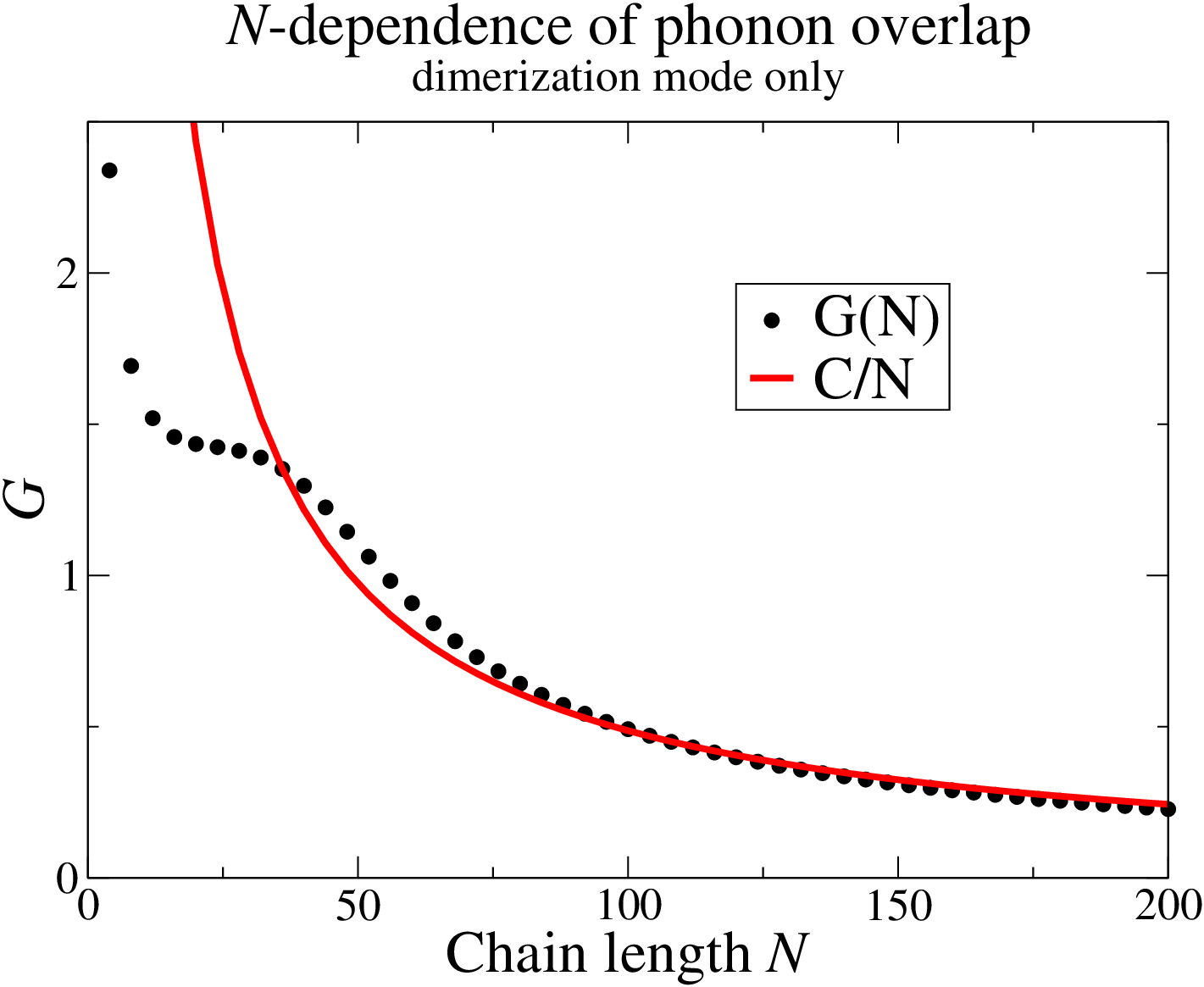}
\end{center}
\caption{Chain length dependence of the phonon overlap $G$, 
considering the dimerization mode only.  For large $N$, we find the overlap 
$\propto 1/N$, as discussed in Appendix A.  For $N>50$ polarons form.}
\label{onemodecomp2}
\end{figure}
\subsection{Beyond dimerization: polaron formation}
\label{subsec:beyond}
Especially in quasi one-dimensional systems, polaron formation upon injection of an
electron becomes an important consideration.  For this section 
\inPaper{of the paper}, we treat the polyacetylene as an N-dimensional
problem, allowing us to study this polaron formation.    

Polarons are quasiparticles formed when an electron is self-trapped in a lattice
deformation.  Polarons are thus electrons dressed in phonons and affect
the conductivity and other properties of a material.

We find that polarons don't form for short chains
(the polaron doesn't fit since its equilibrium length is less than than
the chain length). They do form for longer chains and lead to a finite 
phonon overlap suppression and hence significant
phonon sidebands during electron tunneling.  Because of the
presence of these polarons, these sidebands do not disappear as
$N\rightarrow\infty$.  Similar important polaronic effects should be expected 
for 
other conducting polymers and for 3D systems that form polarons with
substantially suppressed electron mobility.\cite{Emin}  

(This is not to say that real, short polyacetylene rings 
do not distort when charged.  Indeed, benzene ($N=6$) undergoes a dynamic 
Jahn-Teller distortion when a hole is injected,\cite{PerAllPed03} into a shape
which could be viewed as a soliton-antisoliton pair.  The SSH model applied to
benzene dimerizes, however, lifting the electronic degeneracy that drives the
Jahn-Teller distortion.  The quantum phonon fluctuations ignored in the SSH
model (the resonance between the two dimerized benzene states) are necessary to
restore the hexagonal symmetry to neutral benzene, but are too small to destroy
dimerization in long polyenes.  Adding quantum fluctuations to the SSH model
would treat charged benzene as a resonance between two dimerized states; the
more complete model (including bond angle rotations) treats charged benzene as a
dynamical resonance between Jahn-Teller distorted states when phonon quantum
fluctuations are added.)   

\begin{figure}
\begin{center}
\leavevmode
\includegraphics[width=\figWidth]{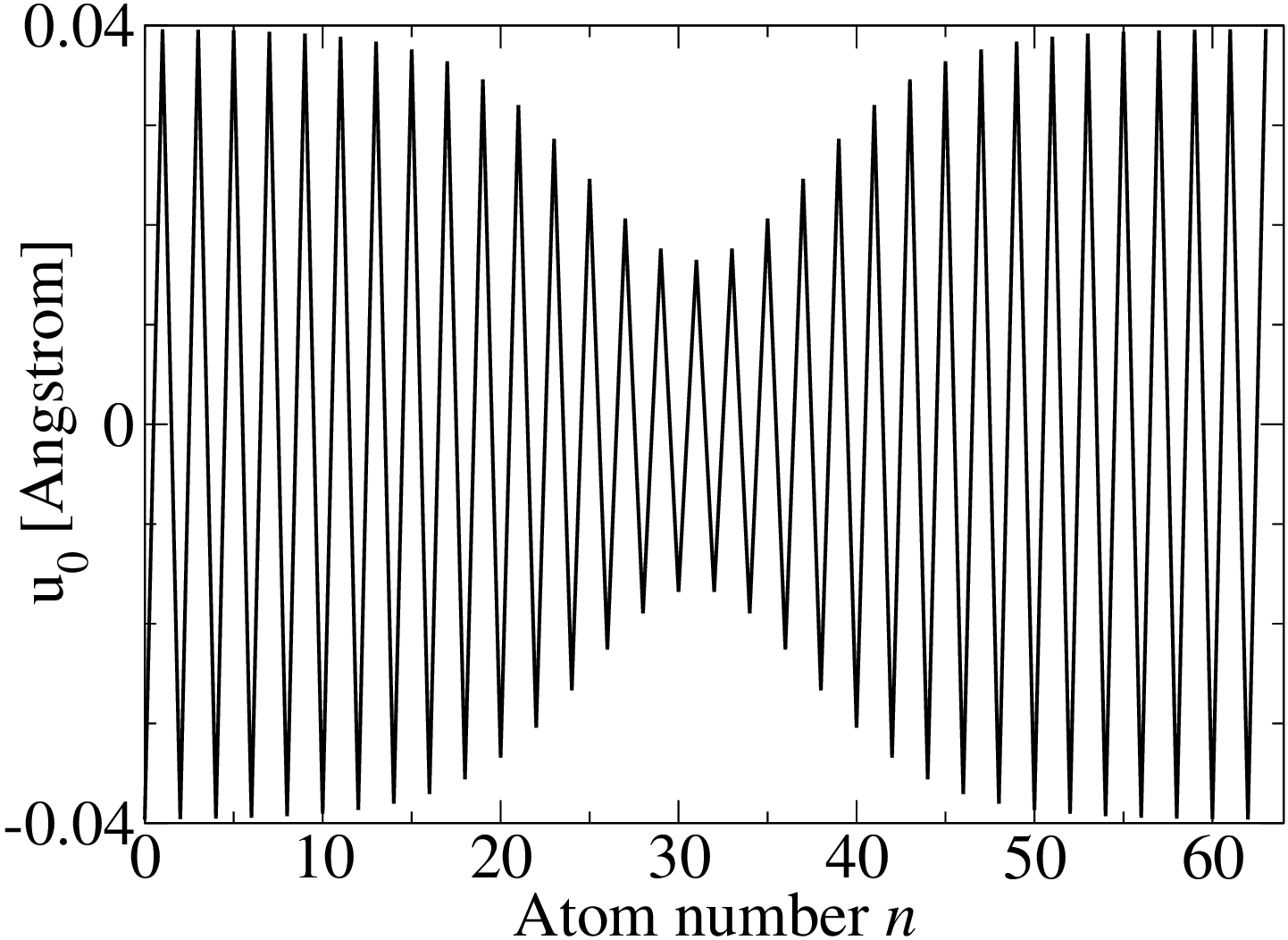}
\end{center}
\caption{Traditional polaron shape (equation~(\ref{start2})). The
shape for molecules of length $\xi\le L$ is closely approximated by one
wavelength of a sinusoid; in this limit the polaron is an optical phonon
with wave vector $k=2\pi/L$.}
\label{start2fig}
\end{figure}

\begin{figure}
\begin{center}
\leavevmode
\includegraphics[width=\figWidth]{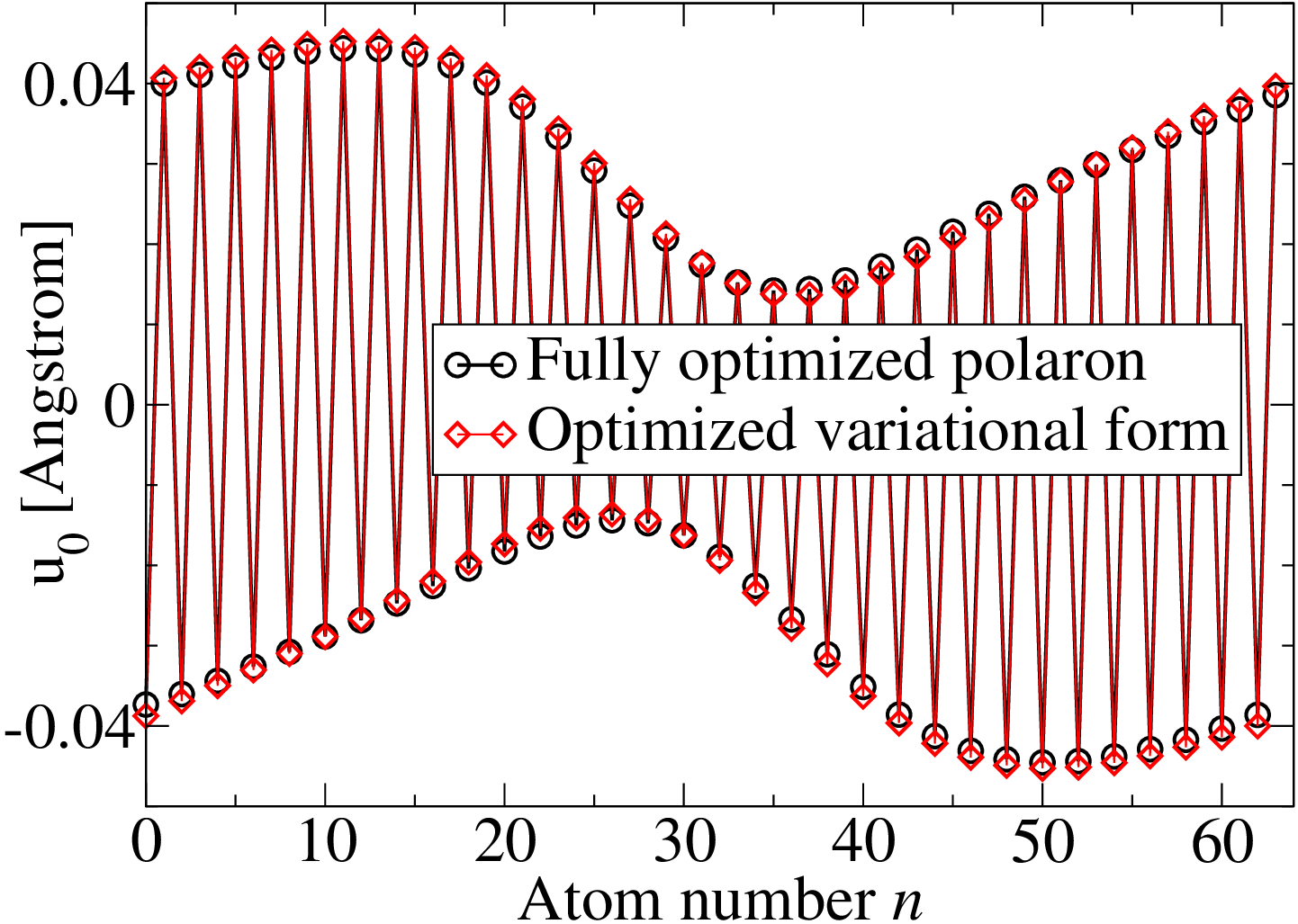}
\end{center}
\caption{When allowing the system to relax fully in the presence 
of an extra electron,
systems of larger than 50 atoms exhibit the polaron.  In addition to the
pinching in the middle, the polaron state also shows a distortion 
of the lattice
at the two ends of the molecule which is due to the periodic 
boundary conditions
and the fact that the chain is contracting.  The diamonds
show the partially relaxed configuration using our functional form 
(eqn~(\ref{functionalForm})); the circles show the fully relaxed
configuration minimizing $N$ coordinates.}
\label{polaron64}
\end{figure}

We first find the minimized geometry of the polyacetylene chain with the
polaron.  The shape of the polaron is traditionally derived from a continuum 
theory, with the
displacement field $\phi_n$ as a sum of two
arctangents (reference~\onlinecite{SSH3}, eqn 4.43):
\begin{eqnarray}
\label{start2}
\phi_n=(-1)^{n}u_{n}&=&u_{0}+{u_{0}\over{\sqrt{2}}}\Big\{\tanh[{(na-b)\over{\sqrt{2}\xi}}]\cr
& &-\tanh[{(na+b)\over{\sqrt{2}\xi}}]\Big\}.
\end{eqnarray}
where $\xi$ is the width of the polaron, $b=.623\,\xi$, $a$ is the lattice
constant,
and $\phi_n$ is the 'staggered' dimerization, a more continuous variable.
Values for these quantities from the literature~\cite{SSH3} are 
$\xi=7.017$ \AA, and $u_{0}\approx 0.04$\AA\ for the charged molecule.  
Figure~\ref{start2fig} shows this traditional continuum ansatz for 
the polaron shape.

However, we found that this traditional form did not describe our relaxed 
configuration, which showed an additional distortion 
(Fig.~\ref{polaron64}).  The polaron in equilibrium has a 
slightly contracted
lattice constant, which pulls in the surrounding atoms.  
By relaxing the total energy within the following functional form
\begin{align}
\label{functionalForm}
(-1)^n u_0
\Bigg( 1 + (A/\sqrt{2}) &
        \Big( \tanh\left(\frac{(n - \frac{N}{2})\,a-b}{\sqrt{2}\,\xi}\right) \cr
&~~~~~~ -\tanh\left(\frac{(n - \frac{N}{2})\,a+b}{\sqrt{2}\,\xi}\right)\Big)\Bigg) \cr
 +~B\,\Bigg( \frac{-2\,(n - \frac{N}{2})}{N}
 &+ \tanh\left(\frac{(n - \frac{N}{2})\,a}{{\sqrt{2}}\, \xi_2}\right) \Bigg)
\end{align}
we reproduce this distortion (Fig.~(\ref{polaron64})).
For long chains (length$=1000$), we find the polaron width parameters
are $\xi=9.5$\AA\ and $b=4.8$\AA, the polaron amplitude increased 
by a factor $A=1.3$,
and the lattice contraction due to the polaron is $B=-0.036$\AA\ spread over a
distance $\xi_2 = 11.3$\AA. (This
corresponds to a shrinkage of the average lattice constant $a$ by a fraction
$X=B/\xi$ over the length $\xi$ of the polaron.) 
We will use the approximate distortion of equation~(\ref{functionalForm}) 
below to estimate the effective mass $m^*$ of the polaron.
\subsection{Phonon overlaps and polarons}
\label{phononoverlapscpolarons}
The general formalism for calculating phonon overlaps and emission 
probabilities is discussed in reference~\onlinecite{C72}. There are certain
subtle modifications needed here in order to deal with the polaron.

When the initial or final electronic state of the system forms a polaron,
one must generalize our formulas to deal with
the almost zero--energy translation mode of the polaron. The hessian
for the charged state of the polyacetylene chain for $N>50$ has two
eigenvalues which are near zero. One is common to all lengths and charge
states and corresponds to the translational symmetry of the molecule; 
the associated eigenvector has all components equal, and this mode can just
be factored out in the calculation (by singular value decomposition or
by hand). The second mode is quite precisely given by the derivative of 
the polaronic deformation (equation~(\ref{start2}),~(\ref{functionalForm}), or 
Fig.~(\ref{polaron64})) with respect to the center $x_0$ of the polaron,
which we will call the polaron translation mode PT:
\begin{equation}
\label{eq:PTMode}
\bheps^{\mathrm{PT}}_{n} = 
	\frac{d\phi_n/dx_0}{\sqrt{\sum_m{\left(d\phi_m/dx_0\right)^2}}}.
\end{equation}
To linear order, this eigenmode translates the polaron sideways, without
overall displacment of the underlying lattice of atoms. Translating the polaron
sideways by one lattice constant brings the polaron to an equivalent 
configuration (with periodic boundary conditions), and hence does not change
the energy. There is a small energy barrier between these equivalent 
configurations (sometimes termed the Peierls barrier), but it is tiny:
exponentially small as the polaron becomes large compared to the lattice 
spacing \cite{joos,sethna}. To an excellent approximation, the polaron is a free particle,
confined to the length of the chain.

Treating the polaron as a particle in a box, we need to find its 
ground--state wavefunction and excitation energies. 
To a linear approximation, shifting the polaron by
$\delta_{PT}$ adds 
$\left(\delta_{PT}/\sqrt{\sum_m{\left(d\phi_m/dx_0\right)^2}}\right)
d\phi_n/dx_0$ 
to the displacement field, which changes the center of the polaron $x_0$
by $\delta_{PT}/\sqrt{\sum_m{\left(d\phi_m/dx_0\right)^2}}$. 
The coordinate $\delta_{PT}$
associated with the polaron translation eigenmode of equation~(\ref{eq:PTMode})
has a kinetic energy $m_{CH}\, {\dot \delta}_{PT}^2/2$. 
Changing
coordinates from $\delta_{PT}$ to $x_0$, the kinetic energy associated with
the polaron translation is:
\begin{eqnarray}
\label{eq:PolaronKinetic}
K_{PT} &=& m_{CH}\, {\dot \delta}_{PT}^2/2\cr 
       &=& \left(m_{CH} \sum_m{\left(d\phi_m/dx_0\right)^2}\right) {\dot x_0}^2/2\cr
       &=& m^* {\dot x_0}^2/2,
\end{eqnarray}
where 
\begin{equation}
\label{eq:EffectiveMass}
       m^* = m_{CH} \sum_m{\left(d\phi_m/dx_0\right)^2}
\end{equation}
and the shift in the polaron center $x-x_0$ given the amplitude $\delta_{PT}$
of the translational eigenvector $\epsilon_{PT}$ is:
\begin{equation}
x-x_0 \sim \delta_{PT} \sqrt{m_{CH}/m^*}.
\end{equation}
The effective mass of the polaron is light.
We calculate the effective mass for a chain of length 
$N=1000$, summing equation (\ref{eq:EffectiveMass}), (with discrete derivatives 
stepping by two) using the analytic expression (equation~\ref{functionalForm})
for the shape of the polaron. We find 
that the polaron mass is 
$1.24\times 10^{-5} m_{CH} \sim 3 m_e$, somewhat larger than the 
value $4.44\times 10^{-5} m_{CH} \sim 1.07 m_e$ found without allowing
for the lattice contraction~\cite{conwell} (equation~\ref{start2} and
figure~\ref{start2fig})
rather than the fully optimized distortion of figure~(\ref{polaron64})).
To ensure that the Born-Oppenheimer approximation is still valid
in our problem, we also calculate the
band-mass of the electron \cite{anm} with the expression
$m^*=\hbar^2/{\partial{E}^2\over{\partial{k}^2}}$, evaluated at
$k={\pi\over{2a}}$ and taking the derivative of the hopping integral.  The band
mass for the electron is $0.15 m_{e}$.  This value still allows for the
Born-Oppenheimer approximation since the light electrons have sufficient time to
relax as the polaron distortion fluctuates.  
The ground--state energy for the polaron center-of-mass coordinate is zero with
periodic boundary conditions.
For a chain of this length, the energy $\hbar^2 k^2/2 m^*$ for the first
excited state $k=2\pi/L$ is 6.29 tHz and the second excited state has energy
56.61 tHz, both in the experimentally interesting
range near the phonon frequencies of Fig.~(\ref{disp}).  

The perhaps more realistic fixed boundary conditions with ground state
$k={\pi\over{L}}$ and excited state $k={{2\pi}\over{L}}$ yields 1.57 THz and
6.29 THz respectively.  For electrons tunneling into the edges of systems
with non-periodic boundary conditions, there could also
be a notable dependence of the electronic matrix element on the polaronic final
state.

To calculate the transition probability into phonon and polaron excited states,
we need the polaronic wave functions.
The one--dimensional wave-function for the polaron center of mass is just
$\psi(x_0)=\frac{1}{\sqrt{L}}\exp(i k_n x_0)$, since we are using periodic
boundary conditions.
The corresponding wave-function in the original phonon-mode basis
$\psi(\delta_{PT})$ is confined to a box of size 
$L \sqrt{\sum_m{\left(d\phi_m/dx_0\right)^2}} = L \sqrt{m^*/m_{CH}}$,
so the corresponding wave-function for the polaron translation mode
is 
\begin{equation}
\psi_n(\delta_{PT}) = \frac{1}{\sqrt{L}}\sqrt[4]{\frac{m_{CH}}{m^*}} 
e^{i k_n \delta_{PT} \sqrt{m_{CH}/m^*}}
\end{equation}
determined to make the square of the wavefunction integrated over the size of
the box equal to one.
The final ground--state wavefunction 
$\Psi^{(2)}_{0,n}$
for the charged polyacetylene molecule 
with a polaron 
is a product of Gaussians for all of the other normal modes of the molecule,
times this one--dimensional wave function,
%\begin{equation}
%\label{minusmodeswf}
%\Psi^{(2)}_{0,n}
%= \frac{1}{\sqrt{L}}\sqrt[4]{\frac{m_{CH}}{m^*}}
%e^{i (2\pi n/L) \sqrt{m_{CH}/m^*} \delta_{PT}} 
%\prod_{\alpha=3}^N
%\sqrt[4]{{m\omega_\alpha\over{\pi \hbar}}}
%	e^{-{{m\omega_\alpha q_\alpha^2}\over{2\hbar}}}
%\end{equation}
where the first subscript of $\Psi_{0,n}$ will be used to denote the number
and type of phonons emitted, and the second will denote the energy level
of the polaronic translation mode.

It is convenient (and necessary for molecules with more than one type of atom)
to change coordinates from positions $\mathbf x$ to 
${\mathbf y} = M^{1/2} ({\mathbf x}-{\mathbf r}_1)$, where $\mathbf r_1$ is the
equilibrium initial configuration. Treating the mass in the SSH model
as a CH mass, and adopting the notation of reference~\cite{C72}, 
the final--state wavefunction becomes:
\begin{eqnarray}
\label{minusmodeswf}
\Psi^{(2)}_{0,n}(\by) &=& \frac{1}{\sqrt{L}}\frac{1}{\sqrt[4]{m^*}}
	N'_2 e^{i \bk_n' \cdot \by}
        e^{-\frac{1}{2\hbar}(\by-\bD)^T \Omega_2 (\by-\bD)} \cr
&=& \frac{1}{\sqrt{L}}\frac{1}{\sqrt[4]{m^*}}
\prod_{\alpha=3}^N\Big\{
\sqrt[4]{{\omega_\alpha\over{\pi \hbar}}}
        e^{i \frac{2\pi n}{L}\frac{1}{\sqrt{m^*}} \by\cdot \bheps_{PT}}\cr
    & & \times 
	e^{-{{\omega_\alpha q_\alpha^2}\over{2\hbar}}}\Big\}
\end{eqnarray}
where 
$\Omega = M^{-\frac{1}{2}} K M^{\frac{1}{2}}$ is the square of the frequency
matrix (and $K$ is the spring constant matrix), 
$\bheps_\alpha$ are the normal modes of $\Omega_2$ with eigenvalues
$\omega_\alpha$ ($\epsilon_{PT}$ being one with eigenvalue near zero), 
$q_\alpha = \by\cdot \bheps_\alpha$,
$\bD = M^{\frac{1}{2}} ({\mathbf r}_1-{\mathbf r}_2)$ is the net atomic
displacement due to the transition, the primed 
$\bk' = M^{-{\frac{1}{2}}} \bk = \frac{2 \pi n}{L\sqrt{m_{CH}}} \bheps_{PT}$
is the rescaled polaron wavevector, and 
\begin{equation}
N'_2 = \prod_{\alpha=3}^N \sqrt[4]{{\omega_\alpha\over{\pi \hbar}}}
\end{equation}
is the normalization of the Gaussian wavefunction with the zero--frequency
polaron translation mode removed.

%The overlap between the uncharged initial state and the charged, 
%polaron--containing final state is then
%  \begin{equation}
%   O_{0,0} = 
%  \frac{1}{\sqrt{L}}\frac{1}{\sqrt[4]{m^*}}
%  {N_{1} N'_{2} \over \bar{N}^{2}} 
%     \exp\left(\tilde{\bD}^T \bar{\Omega} \tilde{\bD}\right)
%	  \exp\left(-\bD^T {\Omega_{2}\over 2} \bD\right)
%  \end{equation}
%where $N_{1}=\sqrt[4]{{det(\Omega_{1})\over{\pi\hbar}}}$, and
%$\tilde{\bD}=(\Omega_{1}+\Omega_{2})^{-1}\Omega_{2}\bD$, from
%reference~\cite{C72}.

The overlap between the uncharged initial state and the charged
polaron--containing final state is then
\begin{eqnarray}
O_{0;0,n} &=& \int d^N y \Psi_0^{(1)*}(\by) \Psi_{0,n}^{(2)}(\by-\bD) \cr
 &=& \int d^N y \Big[
       N_1 e^{-\frac{1}{2\hbar} \by^T \Omega_1 \by}\cr 
    & &  N_2' \frac{1}{\sqrt{L\sqrt{m^*}}}
	 e^{i \bk_n'\cdot \by}
	 e^{-\frac{1}{2\hbar} (\by-\bD)^T \Omega_2 (\by-\bD)}\Big] 
\end{eqnarray}
where the first subscript on $O_{0;0,n}$ represents the initial (ground) state,
the second subscript denotes the phonons emitted in the final state, and
the third is the excitation level of the polaron center--of--mass wavefunction
in the final state.

We rewrite~\cite{C72} this overlap integral as a single
Gaussian with quadratic form $\bar \Omega = (\Omega_1 + \Omega_2)/2$,
centered at $\tilde \bD = {\frac{1}{2}} \bar \Omega^{-1} \Omega_2\cdot \bD$:
  \begin{eqnarray}
  \label{eq:O00n}
  O_{0;0,n} &=& 
  \int d^Ny\Big[e^{i \bk_n' \by} 
   e^{-\frac{1}{\hbar} (\by-\tilde \bD)^T \bar \Omega (\by-\tilde\bD)}\Big]
  \cr
& &\times  \frac{N_1 N_2'}{\sqrt{L\sqrt{m^*}}} 
  e^{-\frac{1}{\hbar} \tilde \bD^T \bar\Omega \tilde \bD}
  e^{-\frac{1}{2\hbar} \bD^T \Omega_2 \bD}\cr
  &=& 
  \int d^N\tilde y\Big[ e^{i \bk_n' (\tilde \by+\tilde \bD)} 
   e^{-\frac{1}{\hbar} {\tilde \by}^T \bar \Omega {\tilde \by}}\Big] \cr
& &\times  \frac{N_1 N_2'}{\sqrt{L\sqrt{m^*}}} 
  e^{-\frac{1}{\hbar} \tilde \bD^T \bar\Omega \tilde \bD}
  e^{-\frac{1}{2\hbar} \bD^T \Omega_2 \bD}\cr
  &=& 
  \int d^N\tilde y\Big[ e^{i \bk_n' \tilde \by} 
   e^{-\frac{1}{\hbar} {\tilde \by}^T \bar \Omega {\tilde \by}} 
  e^{i \bk_n' \tilde \bD}\Big]\cr 
  & & \times\frac{N_1 N_2'}{\sqrt{L\sqrt{m^*}}} 
  e^{-\frac{1}{\hbar} \tilde \bD^T \bar\Omega \tilde \bD}
  e^{-\frac{1}{2\hbar} \bD^T \Omega_2 \bD}.
  \end{eqnarray}

We begin by considering the ground--state to ground--state overlap
$O_{0;0,0}$, leaving the polaron in its ground state $\bk_n=0$. The integral
in the last equation of~(\ref{eq:O00n}) is just 
$\frac{1}{{\bar N}^2}= \frac{1}{\sqrt{\det(\bar \Omega/\pi \hbar)}}$, and
  \begin{equation}
  \label{eq:O000}
  O_{0;0,0} = 
  \frac{N_1 N_2'}{\bar N^2 \sqrt{L\sqrt{m^*}}} 
  e^{-\frac{1}{\hbar} \tilde \bD^T \bar\Omega \tilde \bD}
  e^{-\frac{1}{2\hbar} \bD^T \Omega_2 \bD}.
  \end{equation}

\begin{figure}
\begin{center}
\leavevmode
\includegraphics[width=\figWidth]{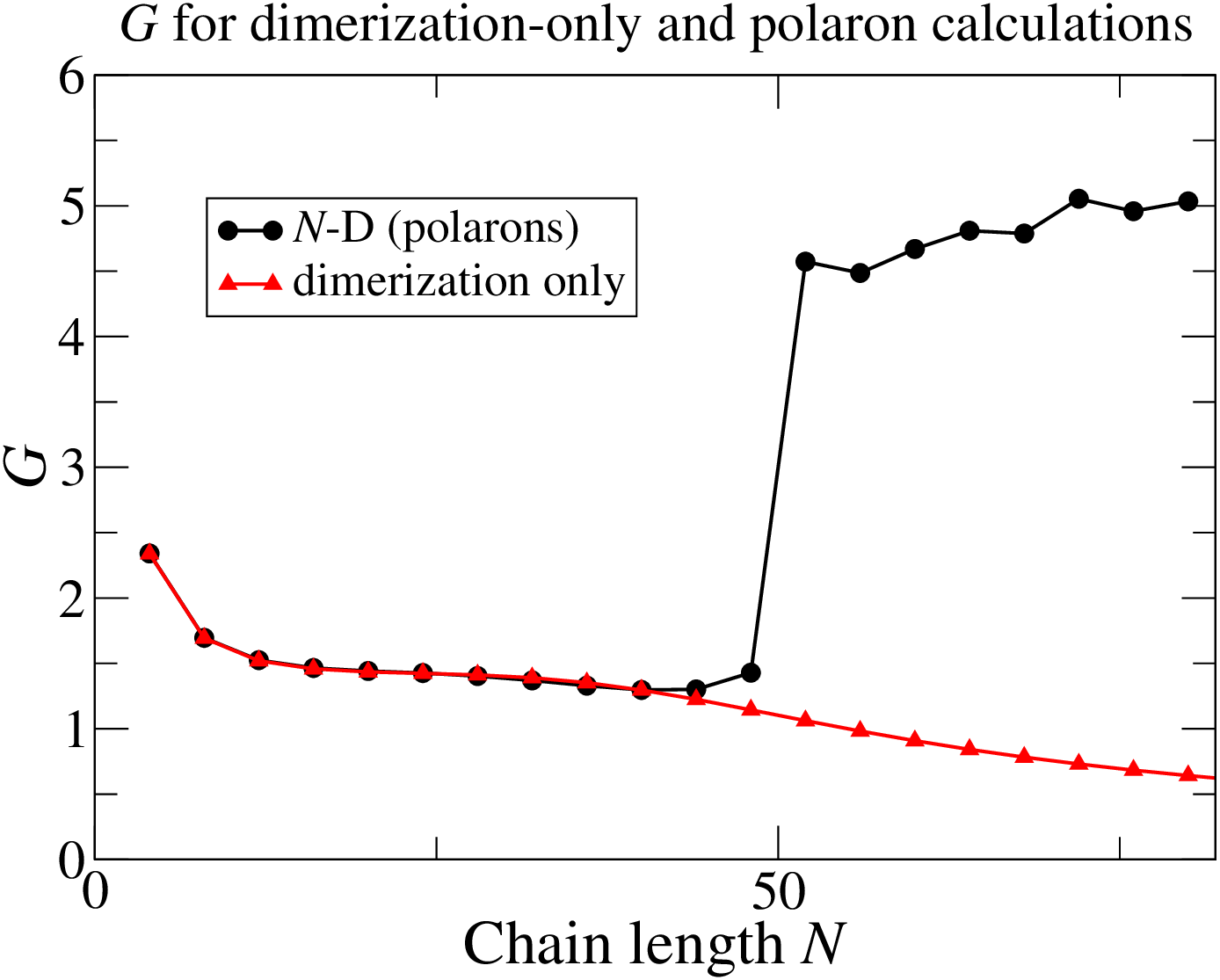}
% used to be manymodes.ps
\end{center}
\caption{$G$ for the case with polaron and without.}
\label{gmodesgraph}
\end{figure}
The total $g$-factor is defined as the suppression of the step height
in the I--V curve due to the ground--state to ground--state
overlap of the atomic wavefunctions, identical to equation~(\ref{G-factor}) save
for the extra polaron component in the final wavefunction:
  \begin{equation}
  G=-\log[|O_{0;0,0}|^2]
  \end{equation}
The presence of the polaron dramatically alters the overlap between the ground
state vibrational wavefunction of the neutral polyacetylene and the ground state
vibrational wavefunction of the charged polyacetylene.  Due to the larger shift
in the geometry, the overlap becomes small and the total g factor becomes large
in comparison to the situation without a polaron.

Figure~(\ref{gmodesgraph}) is a plot of the g factors in the 
dimerization-mode-only 
calculation where no
polaron is allowed, compared to the N-dimensional calculation where a 
polaron is free to form. The graphs coincide for the most part for $N<50$,
where no polaron forms. (The simple calculation involving only the
dimerization mode is approximate even in the absence of a polaron in the
final state because the other phonon modes can emit even numbers of phonons
due to their frequency shifts, even though the configurational shift $\bD$ 
has no component along them.)
The polaron which forms for $N\geq 50$ is associated
with a fixed phonon overlap integral of about 4.5.

\inPaper{
It is natural to ask whether the SSH-model calculations in this paper
can be fleshed out into a quantitative picture, using modern density
functional theory (DFT) calculations for real polyacetylene molecules
(as has been done, for example, for $C_{72}$ and C$_{140}$ 
bi-fullerenes~\cite{ConnieThesis}). 
Early DFT calculations on polyacetylene found no
dimerization after careful convergence in $k$-points~\cite{AshPicKra89};
more recent calculations have shown that the dimerization depends also upon
the basis set and the particular density functional variant
used~\cite{SunKurKer02}. Charged polyacetylene has been 
studied~\cite{SunKurKer02} only 
for effectively rather short systems (one charge per 30 carbons), and while
the total length change was studied no polaronic or soliton deformations
were described. In our initial explorations~\cite{ConnieThesis} we were
able to explore molecules of length $N=20$ using a basis set which
was lacking the diffuse orbitals needed to correctly describe the 
charged state of this conjugated molecule. These calculations showed a
distortion in the charged state primarily associated with the overall
length of the molecule, with little effect on the dimerization.
}

\inPaper{\section{Conclusion}
We have calculated the phonon overlaps for polyacetylene 
using the tight-binding SSH model.
We examined the suppression of the main transition (electronic) by both the
phonons and the polarons, a quasiparticle that is important in
quasi-one-dimensional systems.  The effect of the polaron is to increase
the
bare $G$ for chains that would otherwise have a $G$ that tends
to zero as $N$ increases.  Under certain conditions (when the chain length
$N \geq 50$ and the polaron-in-a-box energy is comparable to the 
phonon energies), the presence of polarons modulate the bare phonon to phonon
transition via the overlap in the same way that phonons modulate the bare
electronic transitions.  
}

\subsection{Acknowledgments}
We would like to thank Jiwoong Park, Abhay Pasupathy, Dan Ralph, 
Paul McEuen, and Geoff Hutchinson for helpful conversations, and Sami Rosenblatt
for his initial work on electron tunneling into polyacetylene. We acknowledge
support from NSF grants DMR-0218475 and CHE-0403806 and from a GAANN
fellowship DOEd P200A030111.

\inThesis{
\chapter*{Appendix A}
\addcontentsline{toc}{chapter}{Appendix A}
}
\inPaper{\section*{Appendix A}}

We can get a qualitative understanding of why phonon emission is more important
for smaller quantum dots by considering how adding an electron to a system with
$N$ atoms changes the bond lengths and hence the phonon overlap. There
are two qualitative cases to consider---the short-wavelength optical phonons
and the long-wavelength acoustic phonons. In the context of polyacetylene,
adding a charge to a molecule with $N$ carbon atoms (ignoring polaron
formation) will change both the dimerization amplitude (an optical
excitation, considered in this manuscript) and the overall length of the
molecular (an acoustic phonon). Charge-induced changes in the length
of {\em trans}-polyacetylene have been studied using density functional
theories~\cite{SunKurKer02}.

For a single phonon mode (and no frequency shift during the transition), 
$G=g$ is given by the square of the atomic displacement
${\bf x}$ between initial and final equilibrium states divided by four times the
square of the zero--point fluctuations $x_0^2$:
  \begin{equation}
  G={{\bf x}^{2}\over{4{\bf x_{0}}^{2}}}={{m\omega}\over{2\hbar}}{\bf x}^2.
  \end{equation}
When an extra electron is added to a quantum dot with N electrons, it usually 
delocalizes over $\sim$ N
bonds. Therefore, its effect on the length of each bond will be diluted, of
order $(\delta a/N)$ where $a$ is the interatomic spacing and $\delta$ is a
constant of order one.  

The optical case is particularly
straightforward: N independent atomic coordinates each shift by $\delta a/N$, so
the Euclidean distance $x^2=N(\delta a/N)^2=a^2\delta^2/N$.  Hence the $g$-factor becomes:
\begin{equation}
\label{g-oned}
G={{\bf x}^{2}\over{4{\bf x}_{0}^{2}}}={\delta^{2}\over{N}}\sim{1\over{N}}.
\end{equation}
Thus implying that optical phonon emission becomes smaller for larger quantum
dots.  We see this also in the explicit SSH calculation summarized in 
figure~(\ref{onemodecomp2}).

% XXX
% This concludes our discussion of the effect of adding an 
% extra electron on the optical mode as a function of increasing length. We found
% that for increasing N, we would expect a corresponding decrease in $g_{\alpha}$
% for the optical mode.   For that
% case, the argument is straightforward.  However, we must also address 
% the effect
% of an extra electron on the acoustic mode of the chain.  We will find that
% $g_{\alpha}$, unlike in the optical phonon case, does not vanish 
% but rather tends 
% towards a constant in the thermodynamic limit.

For acoustic phonons in a D-dimensional quantum dot, we must decompose the
displacement field $u({\bf x})=(\delta/N){\bf x}$ of a dot uniformly rescaled in
size by a factor $\delta/N$ into the orthonormal phonon modes 
$\bheps_{k}=\bhrho e^{i\bk \bx_\bn}/\sqrt{N}$ where 
$\bhrho$ is the
polarization vector (transverse or longitudinal).  
Each mode is excited by an amount $q_k$ where $u=\sum q_k\bheps_k$
with
\begin{eqnarray}
q_k&=&\bheps_k\cdot \bu=\sum_{{{\bf n}}}{{e^{i{\bf k}\cdot \bx_\bn}}\over{\sqrt{N}}}\bhrho\cdot{\bf x}_{{\bf n}}(\delta/N)\cr
& &\approx {1\over{a^D}}\int{d^Dx}{{e^{i{\bf k}\cdot{\bf x}}}\over{\sqrt{N}}}\bhrho\cdot{\bf x}(\delta/N)
\end{eqnarray}
where $\bf n$ labels the lattice sites.
The only contributions are from longitudinal phonons; picking $\bx$, along
${\bf k}$ for a system of size $L$ (with $N=(L/a)^D$) we have:
\begin{eqnarray}
q_k& &\approx {{\delta}\over{N^{3/2}a^D}}\left(\int^{L/2}_{-L/2}dx_1 x_1
e^{ikx_1}\right)L^{D-1}\cr
& &\approx {{\delta}\over{\sqrt{N}}}{1\over{|k|}}.
\end{eqnarray}

The total $g$-factor is the sum of the factors $g_k$ for each normal mode (ignoring for
simplicity the frequency shifts due to the extra electron), 
where we approximate
$w_k\approx ck$ with $c$ the speed of sound:
\begin{eqnarray}
G&=&\sum_k {{m\omega_k}\over{2\hbar}}q_k^2\cr
& &\approx \sum_k{{mck}\over{2\hbar}}{{\delta^2}\over{N}}{1\over{k^2}}.
\end{eqnarray}
We can approximate this sum as an integral, where the density of points in
${\bf k}$-space is $({L\over{2\pi}})^D$, and make a spherical approximation to
the Brillouin zone and the lower cutoff ${1\over{L}}<|k|<{1\over{a}}$:
\begin{eqnarray}
G& &\approx ({{L}\over{2\pi}})^D\left[\int{d^Dk}{{mc}\over{2\hbar
      k}}\right]\left({\delta^2\over{N}}\right)\cr
& &\approx L^D{{\delta^2}\over{N}}\left({{mc}\over{2\hbar}}\right)\int^{1/a}_{1/L}k^{D-2}dk.
\end{eqnarray}
For $D\geq 2$, we have
\begin{eqnarray}
G&=&L^D{{\delta^2}\over{N}}\left({{mc}\over{2\hbar}}\right)\left({1\over{a}}\right)^{D-1}\cr
&=&\left({L\over{a}}\right)^D{\delta^2\over{N}}{{mca}\over{2\hbar}}\cr
&=&\delta^2{{mca}\over{2\hbar}},
\end{eqnarray}
and for $D=1$ we find:
\begin{equation}
G=\delta^2{{mc}\over{2\hbar}}a(\log(L/a))
\end{equation}
We will find in $D=1$ that the electron forms a polaron and is
hence not delocalized, so this logarithmic
divergence is bypassed in practice by polaron formation.

The constant
$\delta^2{{mca}\over{2\hbar}}\approx(a\delta)^2{{m\omega_D}\over{2\hbar}}\approx{{(a\delta)^2}\over{4x_0^2}}$,
where the Debye frequency $\omega_D\approx c/a$.  Hence the acoustic phonons
suppress the quantum tunneling by a finite amount
${{(a\delta)^2}\over{4x_0^2}}$, even for large quantum dots.

% XXX Added
Thus the emission of optical phonons during electronic transitions
becomes unimportant for large systems, while the emission of 
acoustic phonons will continue to suppress the transition by a finite
overlap integral even as $N\to\infty$.

\inThesis{
\chapter*{Appendix B}
\addcontentsline{toc}{chapter}{Appendix B}
}
\inPaper{\section*{Appendix B}}
We now turn to transitions to final states with polarons in excited
translational modes (with polaron wavevectors $k_n$), and with zero
and one phonon emitted.

The integral in the last equation 
of~(\ref{eq:O00n}) is just
the Fourier component of a multidimensional Gaussian at wavevector 
$\bk_n'=\frac{2\pi n}{L\sqrt{m^*}} \bheps_{PT}$. Just as the Fourier
transform of 
$\frac{1}{\sqrt{2\pi}\sigma} \exp\left(-\frac{x^2}{2\sigma^2}\right)
	=\exp(-\frac{k^2 \sigma^2}{2})$, so in many dimensions the Fourier
transform of $\exp(-x^T A x) = (1/\sqrt{\det(A/\pi)}) \exp(-\bk^T A^{-1} \bk/4)$.
Hence
  \begin{eqnarray}
  \label{eq:O00nFinal}
  O_{0;0,n}&=&\frac{1}{\sqrt{\det\left(\frac{\bar {\mathsf{\Omega}}}{\pi \hbar}\right)}} 
	  \exp\left(-\frac{1}{4\hbar} {\bk_n'}^T \bar{\mathsf{\Omega}}^{-1} \bk_n' \right)
  e^{i \bk_n' \tilde \bD} \cr
  & & \times\frac{N_1 N_2'}{\sqrt{L\sqrt{m^*}}} 
  e^{-\frac{1}{\hbar} \tilde \bD^T \bar{\mathsf{\Omega}} \tilde \bD}
  e^{-\frac{1}{2\hbar} \bD^T{{\mathsf{ \Omega}}}_2 \bD} \cr
  &=& 
  \frac{N_1 N_2'}{\bar N^2 \sqrt{L\sqrt{m^*}}} 
  e^{i \bk_n' \tilde \bD} 
  \exp\left(-\frac{1}{4\hbar} {\bk_n'}^T \bar{\mathsf{\Omega}}^{-1} \bk_n' \right)\cr
 & &\times e^{-\frac{1}{\hbar} \tilde \bD^T \bar{\mathsf{\Omega}} \tilde \bD}
  e^{-\frac{1}{2\hbar} \bD^T{\mathsf{ \Omega}}_2 \bD} \cr
  &=& O_{0;0,0}\, e^{i \bk_n' \tilde \bD} 
  \exp\left(-\frac{1}{4\hbar} {\bk_n'}^T \bar{\mathsf{\Omega}}^{-1} \bk_n' \right)
  \end{eqnarray}

This $\bf{k}$-dependence is due to a kind of spatial confinement of the polaron
center of mass in the overlap integral calculation. Indeed, when we
changed coordinates from the polaron center of mass $x$ to the translation
eigenmode $\delta_{PT}$, we made an approximation that the polaron did
not move substantially from its reference position $x_0$. Polaron translations
are given by adding multiples of $d\phi_m/dx_0$ only for shifts $x-x_0$
small compared to the polaron width $\xi$. (A shift by many $\xi$ would
look like the sum of a separated polaron and an antipolaron, not the
derivative of the polaron.) This approximation is valid for our overlap
calculations so long as the `zero-point' Gaussian fluctuations $\xi_0$ of
our overlap integral
along $x$ are small compared to $\xi$. The RMS fluctuations in the polaron
center--of--mass coordinate in the overlap integral are given roughly by
\begin{eqnarray}
\langle (x-x_0)^2\rangle_{\bar{ \mathsf{\Omega}}} 
&=& \frac{1}{m^*} \langle \delta_{PT}^2\rangle 
= \frac{1}{m^*} \langle (\by \cdot \bheps_{PT})^2\rangle\cr 
&=& \frac{1}{m^* \bar N^2} 
  \int d^N y (\by\cdot \bheps_{PT})^2 \cr
& &\times\exp(-\frac{1}{\hbar} \by^T \bar {\mathsf{\Omega}} \by).
\end{eqnarray}
where the expectation is taken in a harmonic potential whose frequency matrix
is $\bar \Omega$, and we ignore the displacement $\bD$ (which does not
shift the center of the polaron). But one can verify that
$\int d\by (\by\cdot {\bf v})^2 \exp(-\by^T C \by) = {\frac{1}{2}} {\bf v}^T C^{-1} {\bf v}$
(say, by going to normal modes of $C$, as in deriving
equation~\ref{brackets} below), so the zero-point polaron fluctuations in our
overlap integral, $\xi_0$, satisfy:
  \begin{equation}
  \label{xi0}
 \xi_0^2=\langle (x-x_0)^2\rangle_{\bar {\mathsf{\Omega}}} 
   = \frac{\hbar}{2m^*} \bheps_{PT}^T \bar{\mathsf{\Omega}}^{-1} \bheps_{PT}. 
 \end{equation}
For our polaron with $N=64$, $\xi_0 = 16.56$\AA$\approx \xi$, so our linearized
approximation is only approximate.  (Comparing, for example,
$u(x_0+\xi_0)-u(x_0)$ in the SSH model with
$u_0+\xi_0{{\partial{\phi_n}}\over{\partial{x_0}}}$ shows a similar distortion,
but with the former centered at $x_0+\xi_0/2$).

Returning to our overlap integral in equation (\ref{eq:O00nFinal}), and noting
again that $\bk_{n}'={{2\pi n}\over{L\sqrt{m^*}}}\bheps_{PT}$, we see 
that the probability of ending in a polaron excited state $n$ with no phonons 
is writeable in terms of this zero-point polaron center fluctuation:
\begin{eqnarray}
\label{eq:P00n}
P_{0;0,n}&=&|O_{0;0,n}|^{2}=|O_{0;0,n}|^{2}e^{-{{\hbar}\over{2}}{\bk_{n}^{'}}^{T}\bar\Omega^{-1}\bk_{n}^{'}}\cr
&=&|O_{0;0,0}|^{2}e^{-{{\hbar}\over{2}}{{4\pi^2 n^2}\over{L^2
	m^*}}\bheps^{T}_{PT}\bar{\Omega}^{-1}\bheps_{PT}}\cr
&=&|O_{0;0,0}|^{2}e^{-4\pi n^{2}({{\xi_{0}}\over{L}})^{2}}.
\end{eqnarray}

And so the g factor associated with this transition is then:
\begin{equation}
g_{0;0,n}=e^{-4\pi^{2}n^{2}({{\xi_{0}}\over{L}})^{2}}=e^{-k_{n}^{2}\xi_{0}^{2}}
\end{equation}
where $k_{n}={{2\pi n}\over{L}}$.
The initial step in the IV curve, where the polyacetylene molecule is left in
its (charged) ground state is thus followed by a series of smaller steps at
energies ${{\hbar^{2}k_{n}^{2}}\over{2m^{*}}}$ and relative heights
  $e^{-k_{n}^{2}\xi_{0}^{2}}$.  

Finally, we turn to the case of one phonon emission combined with excitation of
the polaron translation mode:
\begin{eqnarray}
\label{O01n}
O_{0;1\alpha,n}&=&\int{d^Ny}\Psi_{0}^{(1)*}(\by)\Psi_{1\alpha,n}^{(2)}(\by)\cr
&=&\int{d^Ny}\Big\{N_1e^{-\by^T{{{\mathsf{\Omega}}_{1}}\over{2\hbar}}\by}{{N_{2}'}\over{\sqrt{L\sqrt{m^*}}}}e^{i\bk_n'\cdot\by}\cr
& &\times\sqrt{2\omega_{\alpha}/\hbar}((\by-\bD)\cdot\bheps_{\alpha}^{(2)})e^{-{1\over{2\hbar}}(\by-\bD)^T{\mathsf{\Omega}}_2(\by-\bD)}\Big\}\cr
&=&\Big\{\int{d^N\tilde{y}}e^{i\bk_n'\cdot\tilde{\by}}\tilde{\by}\cdot\bheps_{\alpha}^{(2)}e^{-{1\over{\hbar}}\tilde{\by}^{T}\bar{{\mathsf{\Omega}}}\tilde{\by}}\Big\}\cr
& &\times\Big[\sqrt{2\omega_{\alpha}/\hbar}e^{i\bk_n'\cdot\bD}{{N_1N_2'}\over{\sqrt{L\sqrt{m^*}}}}e^{-{1\over{\hbar}}\tilde{\bD}^T\bar{{\mathsf{\Omega}}}\tilde{\bD}}e^{-\bD^T{{\mathsf{\Omega}}_2\over{2\hbar}}\bD}\Big]\cr
& &-\sqrt{2\omega_{\alpha}/\hbar}((\bD-\tilde{\bD})\cdot\bheps_{\alpha}^{(2)})O_{0;0,n}
\end{eqnarray}

The integral in equation ({\ref{O01n}) can be performed by decomposing
  $\tilde{\by}$ into the normal modes $\bhrho_{\beta}$ of $\bar{\Omega}$ with
  $\tilde{\by}=\sum_{\beta}y_{\beta}\bhrho_{\beta}$:
\begin{eqnarray}
\left\{\right\}&=&\sum_{\beta}\bhrho_{\beta}\cdot\bheps_{\alpha}^{(2)}\int{dy_{\beta}}\,y_{\beta}e^{iy_{\beta}\bk_n'\cdot\bhrho_{\beta}}e^{-{1\over{\hbar}}{y_{\beta}^{2}/{\bar{\omega}_{\beta}}}}\cr
& &\times\prod_{\beta'\neq\beta}\int{dy_{\beta'}}e^{iy_{\beta'}\bk_{n}'\cdot\bhrho_{\beta}}e^{-{1\over{\hbar}}{{y_{\beta'}}^{2}/{\omega_{\beta'}}}}.
\end{eqnarray}
Now $\int{dx}e^{ikx}e^{-Ax^2}=\sqrt{{\pi\over{A}}}e^{-{k^2\over{4A}}}$ and
$\int{dx}e^{ikx}xe^{-Ax^2}={{ik}\over{2A}}\sqrt{{\pi\over{A}}}
e^{-{k^2\over{4A}}}$.
And the overlap becomes:
\begin{eqnarray}
\label{brackets}
\left\{\right\}&=&\sum_{\beta}\bhrho_{\beta}\cdot\bheps_{\alpha}^{(2)}\left({{i\bk_n'\cdot\bhrho_{\beta}}\over{2(\bar{\omega}/\hbar)}}\right)
\prod_{\beta'}\sqrt{{\pi\over{\bar{\omega}_{\beta^{'}}/\hbar}}}e^{-\hbar{{(\bk_{n}'\cdot\bhrho_{\beta'})^{2}}\over{4\bar{\omega}_{\beta}}}}\cr
&=&{{i\hbar}\over{2}}\bk_{n}'^{T}\bar{\Omega}^{-1}\bheps_{\alpha}^{(2)}{1\over{\sqrt{det({{\bar{\Omega}}\over{\pi\hbar}})}}}e^{-\hbar{{\bk_{n}'^{T}\bar{\Omega}^{-1}\bk_{n}^{'}}\over{4}}}.
\end{eqnarray}

So, 
\begin{eqnarray}
O_{0;1\alpha,n}&=&\sqrt{\omega_{\alpha}/\hbar}({{i\hbar}\over{2}}\bk_n'^{T}\bar{\Omega}^{-1}\bheps_{\alpha}^{(2)}-(\bD-\tilde{\bD})\cdot\bheps_{\alpha}^{(2)})e^{i\bk_n'\cdot\bD}\cr
& &\times e^{-{{\hbar}\over{4}}\bk_n'^T\bar{\Omega}^{-1}\bk_n'}
{{N_1N_2'}\over{\bar{N}^{2}\sqrt{L\sqrt{m^*}}}}e^{-{1\over{\hbar}}\tilde{\bD}^T\bar{\Omega}\tilde{\bD}}e^{-{1\over{2\hbar}}\bD^T\Omega_2\bD}\cr
&=&\sqrt{2\omega_{\alpha}/\hbar}({{i\hbar}\over{2}}\bk_n'^{T}\bar{\Omega}^{-1}\bheps_{\alpha}^{(2)}-(\bD-\tilde{\bD})\cdot\bheps_{\alpha}^{(2)})\cr
& & \times e^{i\bk_n'\cdot\bD}e^{-{\hbar\over{4}}\bk_n'^{T}\bar{\Omega}^{-1}\bk_n'}O_{0;0,0}\cr
&=&\sqrt{2\omega_{\alpha}/\hbar}({{i\hbar}\over{2}}\bk_n'^{T}\bar{\Omega}^{-1}\bheps_{\alpha}^{(2)}-(\bD-\tilde{\bD})\cdot\bheps_{\alpha}^{(2)})O_{0;o,n}\cr
&=&(i\sqrt{{{h\omega_{\alpha}}\over{2m^*}}}\bk_n\bheps_{PT}^T\bar{\Omega}^{-1}\bheps_{\alpha}^{(2)}-\sqrt{{2\omega_{\alpha}\over{\hbar}}}(\bD-\tilde{\bD})\cdot\bheps_{\alpha}^{(2)})\cr
& & \times e^{i\bk_n\cdot{\bD}\over{\sqrt{m^*}}}e^{-\bk_n^2\xi^2/2}O_{0;0,0}
\end{eqnarray}
giving us a $g$-factor for the combined excitation of a phonon and a polaron of
\begin{eqnarray}
\label{g1an}
g_{1\alpha,n}&=&e^{-\bk_n^2\xi^2}\Big({{2\omega_{\alpha}}\over{\hbar}}\left((\bD-\tilde{\bD})\cdot\bheps_{\alpha}^{(2)}\right)^{2}\cr
& &+{{\hbar\omega_{\alpha}}\over{2m^*}}\bk_n^2(\bheps_{PT}^T\bar{\Omega}^{-1}\bheps_{\alpha}^{(2)})^{2}\Big).
\end{eqnarray}

The first term is the zero-polaron phonon sideband suppressed by the polaronic
overlap in the same way that in equation (\ref{eq:O000}) the zero-phonon
sideband is
supressed by the phonon overlap.  The second term is due to the change in the
shape of the wavefunction from the polaronic excitation.

\end{document}